# Selective Enhancement in Phonon Scattering leads to High Thermoelectric Figure of Merit in ZnO – Graphene Oxide Core-shell Nanohybrids

Soumya Biswas[1], Saurabh Singh[2], Shubham Singh[1], Shashwata Chattopadhyay[1], K. K. H. De Silva[2], M. Yoshimura[2], J. Mitra[1], Vinayak Kamble[1,*]

[1]School of Physics, Indian Institute of Science Education and Research, Thiruvananthapuram, India 695551.

[2]Toyota Technological Institute, Hisakata 2-12-1, Tempaku, Nagoya 468-8511, Japan.

**Abstract:**

ZnO is a promising candidate as an environment friendly thermoelectric (TE) material. However, the poor TE figure of merit (zT) needs to be addressed to achieve significant TE efficiency for commercial applications. Here we demonstrate that selective enhancement in phonon scattering leads to increase in zT of RGO encapsulated Al-doped ZnO core shell nanohybrids, synthesized via a facile and scalable method. The incorporation of 1 at% Al with 1.5 wt% RGO into ZnO (AGZO) has been found to show significant enhancement in zT (=0.52 at 1100 K) which is an order of magnitude larger compared to that of bare undoped ZnO. Photoluminescence and X-ray photoelectron spectroscopy measurements confirm that RGO encapsulation significantly quenches surface oxygen vacancies in ZnO along with nucleation of new interstitial Zn donor states. Tunneling spectroscopy reveals that the band gap of ~ 3.4 eV for bare ZnO reduces effectively to ~ 0.5 eV upon RGO encapsulation, facilitating charge transport. The electrical conductivity enhancement also benefits from the more than 95% densification achieved, using the spark plasma sintering method, which aids reduction of GO into RGO. The same Al doping and RGO capping synergistically brings about drastic reduction of thermal conductivity, through enhanced phonon-phonon and point defect-phonon scatterings. These opposing effects on electrical and thermal conductivities enhances the power factors as well as the zT value. Overall, a practically viable route for synthesis of oxide - RGO TE material which could find its practical applications for the high-temperature TE power generation.

**Key words:** Al-doped ZnO, Seebeck, thermal conductivity, scattering, defect, oxides.

*Corresponding author email: kbvinayak@iisertvm.ac.in

**INTRODUCTION**

In the present scenario of depleting fossil fuel resources, it is inevitable to switch to renewable avenues for meeting energy demands. One of the safe, economic and easy avenues of energy generation is transformation of waste heat energy to electrical through Thermoelectric (TE) power generation[1, 2]. A significant amount of the heat generated in various physical processes is lost to the surrounding and is under-utilized. Instead with the help of TE devices it could be utilized for producing electrical energy aiding sustainable development[3, 4]. The TE devices shall be of greater interest for their eco-friendly nature along with high power conversion efficiency. However, it also requires environment friendly thermoelectric materials which can transform the waste heat to electric power efficiently[3].

The efficiency of the TE material is a function of a dimensionless quantity known as the *figure-of-merit* (zT), which in the terms of material's physical quantity can be expressed as,[5]

$$zT = \frac{S^2 \sigma T}{\kappa} \qquad \ldots(1)$$

where, T is the absolute temperature, S is the seebeck coefficient, σ is the electrical conductivity and $\kappa$ is the thermal conductivity. The part of denominator, $S^2\sigma$, is referred as power factor and its high value is a reasonable representative of high zT as S and $\sigma$ are correlated to each other via common physical parameter of the material, namely the carrier concentration (n). On the other hand, the denominator, $\kappa$ is the sum of electronic contribution ($\kappa_e$) and the lattice contribution ($\kappa_l$) to heat conduction. Therefore, enhancement in power factor and reduction in thermal conductivity is the key to achieve the high value for the thermoelectric efficiency of a material[6]. A large number of chemical compositions have been explored for high *figure-of-merit* from alloys to oxides and even polymers[6, 7]. While there are a few highly efficient thermoelectric materials like $Bi_2Te_3$, PbTe etc. which shows zT of more than one and the next generation materials having zT of even about two[3, 8]. These are even used in commercial devices however the mere fact that they contain elements like Bi, Te, Sb, Pb etc which are toxic and less earth abundant, demands the scientific community to search for new earth abundant and high TE efficiency materials[9, 10]. In order to overcome the severe problem of environmental hazards of these materials and low abundance, drove a strong push for exploring earth abundant elements and particularly benign oxides. Because oxides are thermally as well as chemically stable, and are easier to synthesize compared to alloys[10].

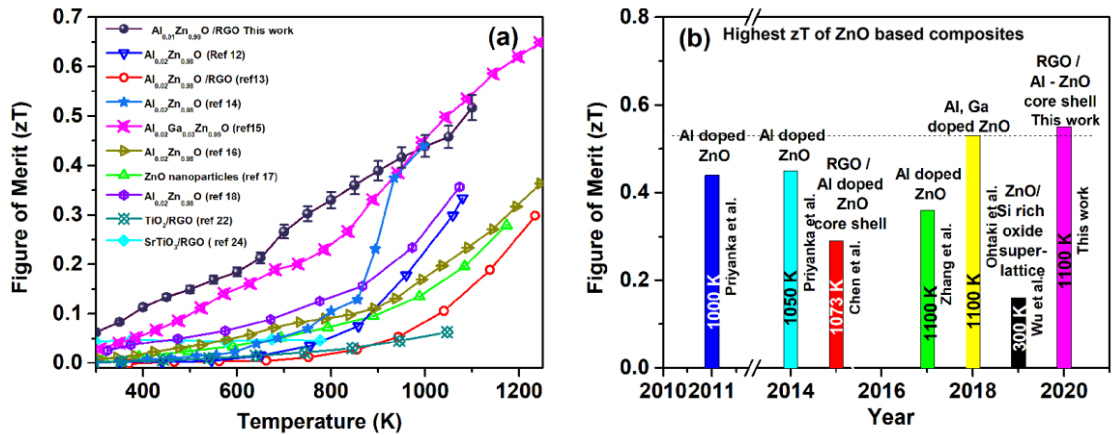

Fig.1. The comparison of the (a) full temperature range and (b) highest thermoelectric figure of merit (zT) of ZnO based systems others reported in literature.

Among the oxides, ZnO, SrTiO$_3$, complex cobaltates with layered structure have been extensively investigated, both experimentally and theoretically[11, 12]. The spin, charge and lattice degrees of freedom allow these materials to tune electronic structure, and provide the favorable condition in suppressing the thermal conductivity by means of phonon engineering to obtain the high *figure-of-merit*[11]. In addition, their high thermal stability, low cost and environmentally benign nature are the strong merits as mentioned earlier. Considering all the suitable parameters which need to be tuned for better TE properties, ZnO is one of the better candidate among oxides materials, which have been extensively investigated[3, 4, 13]. The large band gap, unfavorable phonon dispersion, simple electronic structure of the pure ZnO results in low electrical conductivity and high thermal conductivity, and hence, the poor *figure of merit* (< 0.05). Several researches adopted doping strategies to modify the electronic structure with substitution with Al and Ga at the Zn site in the wurtzite structure, which led to a reasonable improvement in the electrical conductivity and also lowered the thermal conductivity to a certain extent. Enhancing the electrical conductivity by introducing the excess carriers is favorable for large power factor, while, in order to achieve large zT of this material, thermal conductivity value is required to be suppressed via phonon engineering[14]. These dopants also act as scattering centers for the phonons to reduce the thermal conductivity. Hitherto, there are a number of reports showing enhancement in the zT of ZnO by the microstructural engineering to particularly reduce the thermal conductivity while retaining a moderate power factors ($S^2\sigma$).

Figure 1 shows the comparison of the maximum zT values reported in the past several years for doped ZnO systems and that of oxide RGO composites. The trivalent donor impurities like Al, Ga, In at the Zn site are found to enhance the electrical conductivity by increasing carrier concentration in the conduction band[15, 16, 17]. This also reduces the thermal conductivity by enhancing the point defect scattering[18]. Thus, microstructure tuning and doping trivalent ions has been a successful approach to get a high zT value[3, 13, 15, 16, 18, 19]. For instance, Zhang et al.[20] prepared a mixture of micro and nanostructured ZnO which led to a high zT of 0.36 at 1073 K. While, Ohtaki et al.[18] synthesized ZnO with the co-doping of Al as well as Ga which produced the so far best zT of 0.5 at 1100 K as shown in Fig 1. The details of the various reports on comparison of zT of ZnO based thermoelectrics are represented in Fig 1(a) and 1(b). Recently, Jood et al.[16] prepared the Al-doped ZnO using rapid microwave method and achieved a maximum zT of 0.44 at 1000 K.

The recent breakthrough in oxide semiconductors has emerged with composites of graphene and other core-shell oxide particles due to its conducting nature[15, 21, 22, 23]. These reports show that graphene or Reduced Graphene oxide (RGO) when capped on to metal oxide particles enhance the zT drastically. However, the extent of RGO reduction and the nature of RGO oxide interface largely decides the thermoelectric transport of these composites[15, 21, 24, 25, 26]. A one step synthesis process adopted by *Chen et al.*[15] led to Al-doped ZnO-RGO nanocomposite showing a maximum zT of 0.28 at 1173 K. The presence of even a thin layer of RGO interface changes the optical and electronic properties of ZnO drastically as demonstrated by Son et.al[27]. Here, it is important to note that, both the optimization of power factor, and suppression of thermal conductivity is very crucial. Thus, the choice of additives, together with amount of RGO is responsible for phonon scattering. These are taken care by choosing best composition and synthesis condition to show the synergic effect through a delicate balance between the two.

In this work, we show that through systematic investigation the high figure of merit obtained so far in Al doped ZnO and RGO composite results from the nature of interface between the two and particularly the electron and phonon transmission across this interface. The hybrid material is synthesized by a facile solvothermal method followed by Spark Plasma Sintering (SPS). The hybrid structure of RGO encapsulated Al doped ZnO (AZO) quantum dots is found to very effective to suppress the thermal conductivity via phonon scattering and provide large power factor at the same time leading to the large zT value. The extensive investigations reveal that smaller initial particle size and non-equilibrium conditions of SPS lead a better reduction and coverage of RGO significantly affecting the thermoelectric transport of AZO quantum dots leading to significantly higher figure of merit in lower temperature and a maximum value of ~0.52 at 1100 K.

## MATERIALS AND METHODS

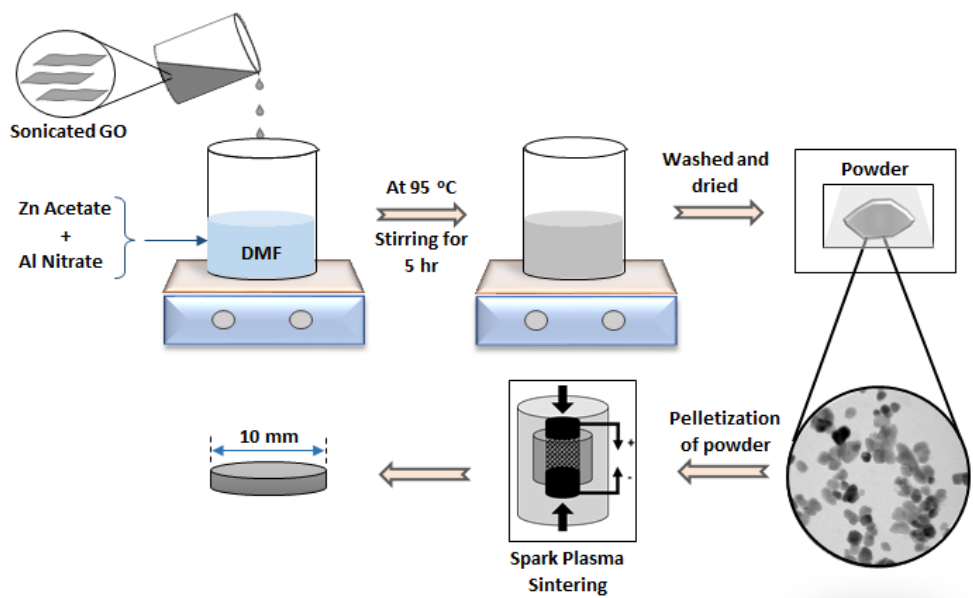

**Figure 2.** Schematic illustration of synthesis of RGO encapsulated Al doped ZnO quantum dots powder by solvothermal method and the pelletization by the SPS.

Figure 2 shows the schematic diagram of the synthesis procedure. The sample codes and respective composition information is summarized in Table 1. The undoped ZnO and ZnO-Graphene core-shell QDs (GZO) were prepared using method reported in the literature[27]. The same has been modified to obtain the 1% Al doped variants of bare (AZO) and RGO encapsulated (AGZO) QDs.

The graphene oxide (GO) used in the study was prepared using modified Hummers method[28]. In a typical synthesis, 5 mg of GO was dispersed in 40 ml of *N-N* Dimethyl Formamide (DMF) under sonication. Simultaneously, 0.92 gm of zinc acetate dihydrate ($Zn(CH_3COO)_2 \cdot 2H_2O$) was dissolved in 200 ml of DMF. For 1 at% Al doping, 19 mg of aluminium nitrate nonahydrate ($Al(NO_3)_3 \cdot 9H_2O$) was added into this solution. The GO solution in DMF was added drop-wise to Zn acetate solution under constant stirring. This solution was maintained at 95 °C and subsequently left for 5 hours under constant stirring. This elided grey white colored suspension was recovered by repeated washing with ethanol and followed by drying at 55 °C in the vacuum overnight. For preparing without RGO, bare ZnO (ZnO) and 1% Al-doped ZnO (AZO) particles, exactly the same method was followed, except the addition of sonicated GO solution.

Table1. The details of sample code, composition and their sizes as well as densities.

| Sample Code | Composition | Crystallite size from XRD (nm) | Average Crystallite Size from TEM (nm) | Density after SPS (gm /cc) | % Density wrt theoretical density of ZnO | EPMA Composition |
|---|---|---|---|---|---|---|
| **ZO** | Only ZnO | 25 ± 5 | 25 ± 5 | **5.31** | **94.7** | $ZnO_{0.95}$ |
| **AZO** | 1 at% Al doped ZnO | 20 ± 5 | 20 ± 5 | **5.22** | **93.0** | $Zn_{0.9904}Al_{0.0096}O_{0.94}$ |
| **GZO** | 2 wt% RGO encapsulated ZnO | 20 ± 5 | 20 ± 5 | **5.23** | **93.2** | $ZnO_{0.98}$ + (0.98 wt% C) |
| **AGZO** | 2 wt% RGO encapsulated and 1 at% Al doped ZnO | 17 ± 5 | 20 ± 5 | **5.19** | **92.5** | $Zn_{0.986}Al_{0.014}O_{.987}$ + (1.05 wt% C) |

The initial characterizations were performed on the as prepared powders of QDs followed by the same on sintered sample. This includes X-ray diffraction, Transmission electron microscopy for structural and morphological characterizations. The powders were densified in pellet of 10 mm diameter using Spark Plasma Sintering (SPS) under inert (Ar) atmosphere using a graphite die. The pressure employed was 50 MPa and temperature achieved using plasma heating was about 1273 K. The as-prepared and sintered samples were examined using

Philips Panalytical X-pert Pro X-ray diffractometer having Cu $K_\alpha$ X-ray source of 0.15418 nm in the range of 5 - 90 degrees and step size of 0.017 degrees. Further, the Raman spectra were recorded using 532 nm laser source on Horiba Jobin Vyon LabRam HR Raman and Renishaw – InVia Raman spectroscope by exciting with 532 nm laser at room temperature. The High resolution transmission electron microscopy was undertaken on FEI 300 keV Microscope equipped with an EDS analyzer. The morphology and mapping analysis of sintered pellets are examined by scanning electron microscope (SEM) using a JEOL JSM-6330F and energy-dispersive x-ray spectroscopy (EDS) using a JED-2140GS with the accelerating voltage of 20 kV respectively. The composition of the samples was analysed by electron probe micro analysis (EPMA) using a JEOL JXL-8230. A Scanning Tunnelling Microscopy (STM) and Scanning Tunnelling Spectroscopy (STS) analysis of the samples were done using R9 SPM control system RHK technology UHV (Ultra High Vacuum STM) using a sharp Pt-IR wire as tip.

The transport measurements were performed on the sintered samples. The electrical resistivity is determined by the four probe method and the seebeck measurement was done by an in-house build instrument[29]. The sample density measurements were done by using the Archimedes principle. The thermal conductivity measurements were done using laser flash analysis method (NETZSCH, LFA 457). The Photoluminescence (PL) spectra were recorded using Fluorolog-3, Horiba Jobin-Vyon spectrofluorometer. The X-ray Photoelectron Spectroscopy (XPS) was done by using Omicron ESCA 2SR XPS and ULVAC-PHI 5000 Versa Probe II system equipped with Mg $K_\alpha$ 1253.6 eV and Al $K_\alpha$ 1487 eV photon energy respectively, at room temperature. The binding energies reported here are having an accuracy of ±0.3 eV. The system is equipped with auto-charge neutralizer. For preparing the sample for XPS, the powder was first dispersed in the ethanol and sonicated for about 2 minutes. Subsequently, the dispersed solution was drop-casted on the Si/SiO$_2$ substrate.

**RESULTS AND DISCUSSION**

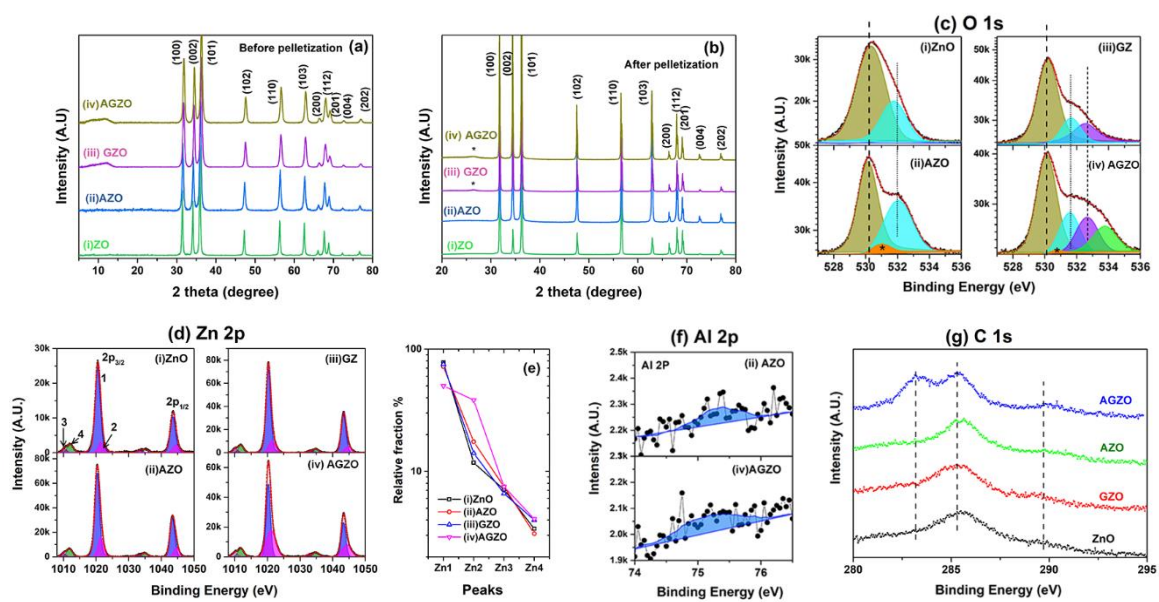

*Figure 3. The XRD patterns of the bare as well as RGO encapsulated ZnO and Al doped ZnO samples (a) before and (b) after pelletization using spark plasma sintering. The X-ray photoelectron spectra comparison for (i)ZO, (ii) AZO, (iii) GZO and (iv) AGZO in (c) O 1s, (d) Zn 2p, (e) comparison of relative Zn 2p peaks, (f) Al 2p and (g) C 1s core levels.*

Figure 3 describes the XRD pattern of the as prepared as well as densified nanoparticles. The patterns have been indexed corresponding to a hexagonal wurtzite structure (space group P6$_3$mc, JCPDS No. 36-1451) with the lattice parameter *a* = 3.250 Å and *c* = 5.214 Å for bare ZnO. Further, the lattice parameters were calculated for undoped ZnO as **a** = 3.250 Å, **c** = 5.207 Å. The GZO and AGZO powder samples shows a broad peak at low Bragg angles, around 12º (Fig 3(a)) which confirms the presence of the GO. After SPS led pelletization, the XRD was performed again on the pellet samples to see changes in crystallinity, if any. Surprisingly, a peak is observed at 25º due to RGO (marked by asterisk* in Fig 3(b)). The claim is further strengthened by the fact that this peak is not observed for samples without RGO i.e. ZO and AZO.

The minimum crystallite size obtained from the powder sample is around 15 - 20 nm as shown in Table 1. The broadening of the peaks in the AZO could be due to Al doping, which might suppress grain growth, resulting in smaller crystallites. (The full range comparison of XRD patterns of ZnO with and without Al dopant can be found in Fig. S1 in Supporting information section). No notable peak shift was observed in Al doped samples, maybe due to low doping concentration and broad peak profiles. Given the comparable ionic radii of Al and Zn, it is highly likely that Al may substitute at Zn site in low dopant concentrations[43]. Several reports show that the solubility limit of $Al^{3+}$ into $Zn^{2+}$ sites in ZnO, is about 2 at% or more. Thus, it is possible that in this case Al substitutes at Zn site. As seen in Fig. 3(a and b), no additional peaks were observed before or after pelletization within the resolution of the data, suggesting that the sample have purely single phase.

The XPS investigations were performed on the drop-casted powder samples prepared as mentioned earlier. Fig 3(c-g) shows the XPS analysis of the composites. As shown in Fig 3(c), the deconvolution of the O 1s spectra was done on the basis of different chemical states of the surface. In all 4 spectra, the main O 1s peak was found to be nearly 530.2 eV. It is ascribed to the O-Zn bonding in ZnO tetrahedra in the ZnO lattice. The peak around 532 ± 0.3 eV has been ascribed to the presence of oxygen vacancy ($V_o$) defects in the ZnO lattice[30]. This peak shows a significant variation in different samples as seen in Fig 3(c). In case of the RGO incorporated samples, i.e. GZO and AGZO, an additional peak seen in the spectra at nearly 533 eV. The presence of this peak in only RGO incorporated sample points towards its origin from O - C bond in the RGO sheets. Additional peaks on far high binding energies like ~534 eV are usually ascribed to the chemisorption or the complex organic bonding with the oxygen. While, for Al doped sample another additional peak appeared (although in small intensity, and denoted by the asterisk*) around ~531 eV. Again, as it is only found in Al doped samples, confirms its correlation with Al in the Zn site. The change in local chemical environment of oxygen ion, surrounded by Zn and Al both may result in a small difference in the binding energy of the outgoing photoelectrons[31].

The study of the Zn 2p core level spectra are very crucial for these composites for the following reason. In the Fig. 3(d) the Zn 2p spectra of the composites are presented. As expected, the two main peaks correspond to the spin orbit splitting of the Zn 2p i.e. Zn $2p_{3/2}$ and Zn $2p_{1/2}$ appears at 1021.1 eV and 1044.2 eV respectively. Ideally, Zn assumes only one oxidation state and the only equivalent lattice site in ZnO, hence no other chemical shift etc. is expected in this region. However, a few peaks are observed including an immediate higher binding energy peak for each of the Zn $2p_{3/2}$ and Zn $2p_{1/2}$ around 1022.2 eV and 1045 eV respectively. This is most likely due to the surface OH-Zn-OH bonding, which can make the apparent high binding energy for surface Zn ions[32]. Besides, two small peaks are found around 1012 eV and 1036 eV which is very low in energy compared to the main Zn peaks (about 10 eV lower)[33]. However, these appeared consistently for this sample and their area ratios etc, satisfied the spin-orbit criteria for a *p* core level. These could possibly be due to inherent defects of Zn ions occupying the interstitial sites ($Zn_i$). Their relative fraction with respect to main Zn $2p_{3/2}$ peak was calculated for each sample and the quantitative analysis is shown in Fig. 3(e). Notably, it is the least in bare ZnO, and is significantly high for RGO-Al containing composite than that of others.

The Al 2p core level region is shown in Fig 3(f). In the Al 2p spectra, there is a very small change in the signal intensity compared to the background, yet visible on comparison with undoped samples. (See Supporting Information Fig S2 (b) for survey spectra and the Al 2s spectra comparison with the undoped samples.) This could be due to very small doping of Al i.e. nominal 1% and the small coverage of sample in drop casting. Similarly, in the C1s spectrum the peaks appeared due to the C-C bond at about 283 - 284 eV[31]. A high binding energy peak is seen prominently in doped samples. However, the overall signal is low and in distinguishable if it is from RGO or surface adsorbed carbonaceous things, which is again due to the small relative fraction of RGO. Besides, no significant difference is found in these spectra (Fig. 3(g)).

The XPS was also performed on the polished (to remove the surface impurities) sintered pellet samples. (In the Supporting information section, the spectra are given in Fig S3). Due to the close packing and high densification the observed peaks at high binding energy in the powder samples for Zn 2p spectra (1012 eV and 1036 eV) is clearly diminished in the case of pellet samples. In the O 1s spectra recorded on the pellets is nearly similar for all the samples except AGZO. The main peak due to O-Zn bonding is clearly visible and a small shoulder peak around 533 eV appears for the oxygen vacancies as mentioned earlier. However, in the case of the AGZO sample this peak is more prominent than other three samples. In the densified pellets significant presence of the C-C bond around 285 eV is visible which is visibly different with that of RGO lacking samples (ZnO and AZO) having dominant peak at ~286 eV. In the GZO and AGZO pellets the presence of RGO is confirmed by the broadening of C1s spectra due the sp2 and sp3 hybridized C-C bond around 284 eV and 285 eV respectively[31].

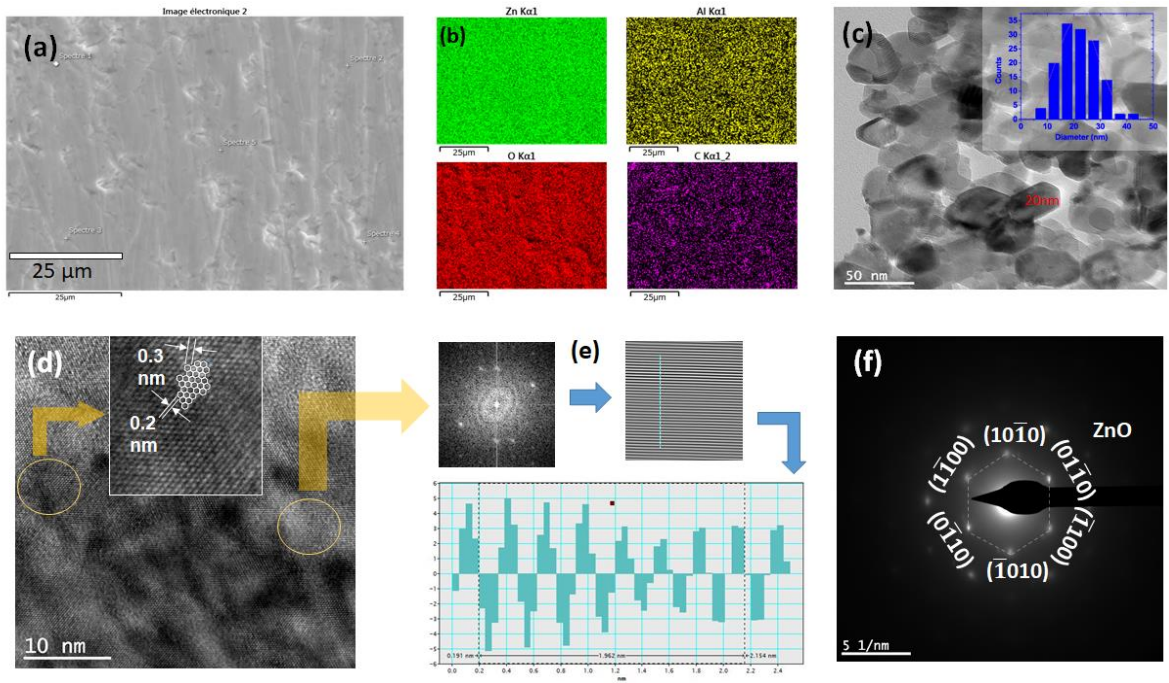

*Figure 4.* *(a) SEM of the AGZO pellet (b) EDS mapping (c)HRTEM of the same sample (the inset: particle size distribution) (d)Magnified TEM image of a particle (inset: hexagonal ordering matching Graphene interplanar spacing) (e)FFT analysis of the selected portion (f) SAED image of the ZnO powder.*

The densities of the SPS sintered samples were measured using Archimedes method. As seen in Table 1, fairly dense sintered bodies are obtained having densities ~95% of the theoretical density of ZnO (5.61 gm/cc.). The bare ZnO is ~95% dense, this is of prime importance for having high electrical conductivity. The density of Al doped and composite is slightly lower due to the presence of the constituent phases. Fig. 4(a) shows the SEM image of the densified AGZO pellet having a texture. The compact nature of the sample confirms no visible porosity, which is also consistent with the measured and calculated density. The EDS mapping was performed to study the compositional uniformity of *Al* dopant and RGO additive. Fig 4(b) shows the elemental map of *Zn, O, Al* and *C*. It may be clearly observed that there is a uniform distribution of Al and C in the composite. Before sintering, the powder sample of QDs were analyzed using the transmission electron microscopy (TEM). The high resolution TEM images are seen in Fig 4(c) which shows well separated QDs. The size distribution of the particles was plotted and the result yields an average size of 20 ± 5 nm. (see more TEM images in Fig S4 in Supporting information section) Fig. 3(d) shows the HRTEM of ZnO reflections forming Moiré fringes. The reflections were identified to have hexagonal pattern, corresponding to honeycomb lattice of Graphene (RGO) encapsulating the ZnO particles. The Fast Fourier Transform (FFT) analysis was performed to determine the interplanar spacing as shown in Fig 4(e). The value obtained, 0.28 nm was found to be in good agreement with the lattice spacing of ZnO for (100) reflections. After palletization, the TEM images were again recorded and the sample shows larger crystallites with large agglomeration (Fig.3(f)). The HRTEM shows hexagonal pattern which when analyzed shows the lattice spacing of graphene like honeycomb structure in most regions and intermittently lattice spacing of ZnO (100) reflection. The Selected Area Electron Diffraction (SAED) pattern (shown in Fig 3(f)) was obtained and indexed. It shows the hexagonal symmetry of ZnO lattice along <002> zone axis. The spots are indexed accordingly. The EPMA was done on the samples to quantify the exact composition of the samples. It is given in the Table 1. (For additional quantification data of EPMA please see Figure S5 and Table S2, S3 in supporting information section)

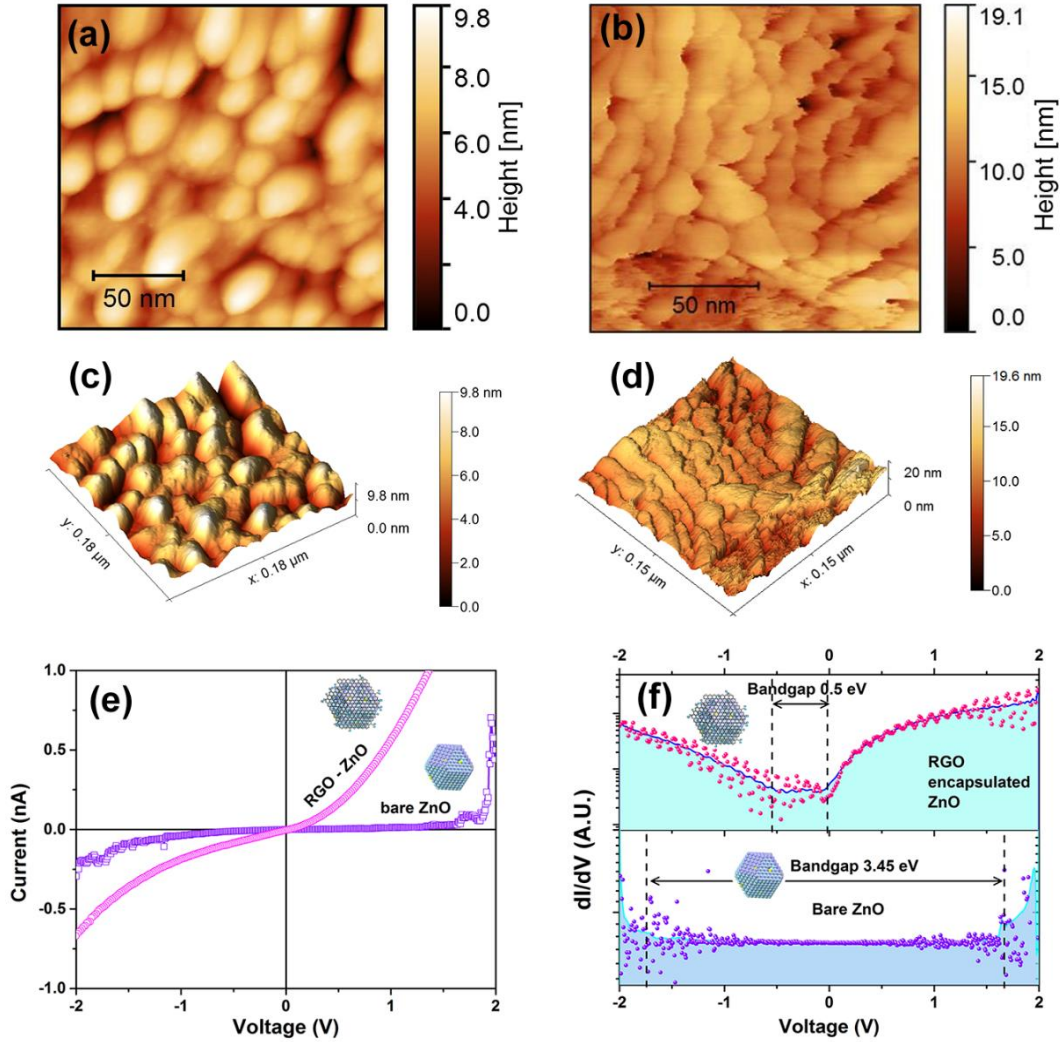

*Figure 5. Scanning Tunnelling Microscope topographical images and 3D reconstruction of (a)& (c) bare ZnO nanoparticles and (b)&(d) RGO encapsulated ZnO nanoparticles (GZO) on the gold coated glass substrate. (e)I-V characteristics and (f) dI/dV Vs V plot showing density of states and the measured energy gap of bare ZnO and GZO as prepared nanoparticle samples.*

To confirm the microstructure of RGO shell on ZnO nanoparticle cores in the GZO sample, Scanning Tunneling Microscopy (STM) and Scanning Tunneling Spectroscopy (STS) were performed on the bare ZnO as well as the GZO powder. The topography of the samples was obtained at fixed current mode. The as prepared nanoparticles with and without RGO capping were drop casted on a glass substrate having a gold film. The topographical image of ZnO sample (Figure 5(a,c)) was recorded at bias voltage 896 mV and tunneling current 300pA. Similarly, for the GZO sample it was recorded (Fig. 5(b,d)) at 846 mV keeping the tunneling current fixed at 400 pA. Both exhibit particle size of 20-30 nm which is in good accordance with that of the TEM images, however the RGO encapsulated ZnO sample showed smaller surface roughness and lesser voids (gaps) between individual nanoparticles, which likely improves inter-particle electrical contact and reduces resistance. Subsequently, on performing the STS on the both samples revealed the distinctly different nature of the two samples as seen in Fig 5 (e) and (f). As this experiment mainly deals with the surface properties of the samples, it is expected to get the electrical signature of the capping, if any. This is manifested by a significantly smaller energy gap evidenced for the GZO sample than that of the case of bare ZnO nanoparticles, due to the presence of RGO encapsulation. The STS analysis was carried out to observe the I-V as well as the dI/dV characteristics of both the samples. The electronic band-gaps was estimated using a lock-in detection by measuring the dI/dV signal from the tunneling current. A reference signal of frequency of 1 KHz and amplitude 30mV was applied. The observed I-V and corresponding dI/dV vs V data of both the samples are given in Fig. 5(e) and Fig. 5(f) respectively. The RGO encapsulation results in a drastically lower band gap ($E_g$) value of 0.56 eV[34] viz-a-viz ~3.4 eV of ZnO. The data further points towards the fact that the presence of RGO shell provides a low resistance surface path for the current bypassing the ZnO core. Thus, in a stack of such particles, the RGO chain formed between the consecutive ZnO nano particles is likely to manifest the band-gap of RGO only. Thus this lower effective band gap is comparable with previously observed band gaps of RGO confirming its presence as a capped layer over the ZnO nanoparticles. In order to verify this, the same experiment was also performed bare ZnO nano-particles. As seen in Fig 5(f), the measured band gap value for bare ZnO around 3.4 eV which is quiet close to that of expected value. The line profile on the images were measured to confirm the particle size. The same can be found in Fig S6 in the supporting information section. These are of good agreement with the calculated particle size from the TEM images.

Figure 6(a) and 6(b) presents the Raman spectra of the as-prepared samples. The Raman spectrum of ZnO shows a conspicuous signal of the $E_2$ (high)[35] mode at 438 cm$^{-1}$. Further, the peak at nearly 340 cm$^{-1}$ corresponds to 3$E_2$ (high) – $E_2$ (low), and $E_1$(TO) mode around 380 cm$^{-1}$ for all the composites. (See supporting information Fig S7 for the details of stretching vibrations). However, in the Al doped samples (AZO, AGZO) the relative intensity of the $E_2$ (high) mode is substantially suppressed and a broad peak is observed at around 550 cm$^{-1}$ instead. This peak appears ($A_1$ (low optical) or $E_1$ (low optical)) due to the breaking of translational symmetry of the ZnO due to the Al dopant. In other words, due to the substitutional Al impurity defect states in the doped ZnO[36]. The inclusion of RGO is confirmed by the D and G peaks at 1330 cm$^{-1}$ and 1580 cm$^{-1}$ respectively. A rather broad FWHM of D and G peaks along with significantly high D peak intensity is a signature of RGO[31]. However, no 2D peak could be seen in the spectra of as synthesized powders. This may be due to poor exfoliation of GO (a few layers). These spectra confirm the doping and the RGO inclusion in the composites. The Raman spectra was also taken on the sintered pellet samples. The same are shown in Fig. 6(c) and 6(d). (for uniformity check full range Raman spectra for all sintered samples across sample surface were taken and are shown in Fig S8 in supporting information). Notably, there are significant changes in the spectra after pelletization. Although all four samples consist of the $E_2$ high mode at 438 cm$^{-1}$, the intensity of this peak is drastically lower in case of the Al-doped samples. (See Fig S8 in supporting information form magnified view of ZnO region of Raman spectra). The peak around 1120 cm$^{-1}$ is also clearly visible for all the samples. This is ascribed to the $A_1$(TO)+$E_1$(TO)+$E_{2L}$ mode[36]. Nevertheless, the most notable change has occurred in case RGO contribution of the spectra. Two peaks of 2D and D+G modes are now conspicuously visible at nearly 2700 and 2900 cm$^{-1}$. These peaks denote the exfoliation of RGO during the process. These results are in very good agreement with those of the XRD and microscopy.

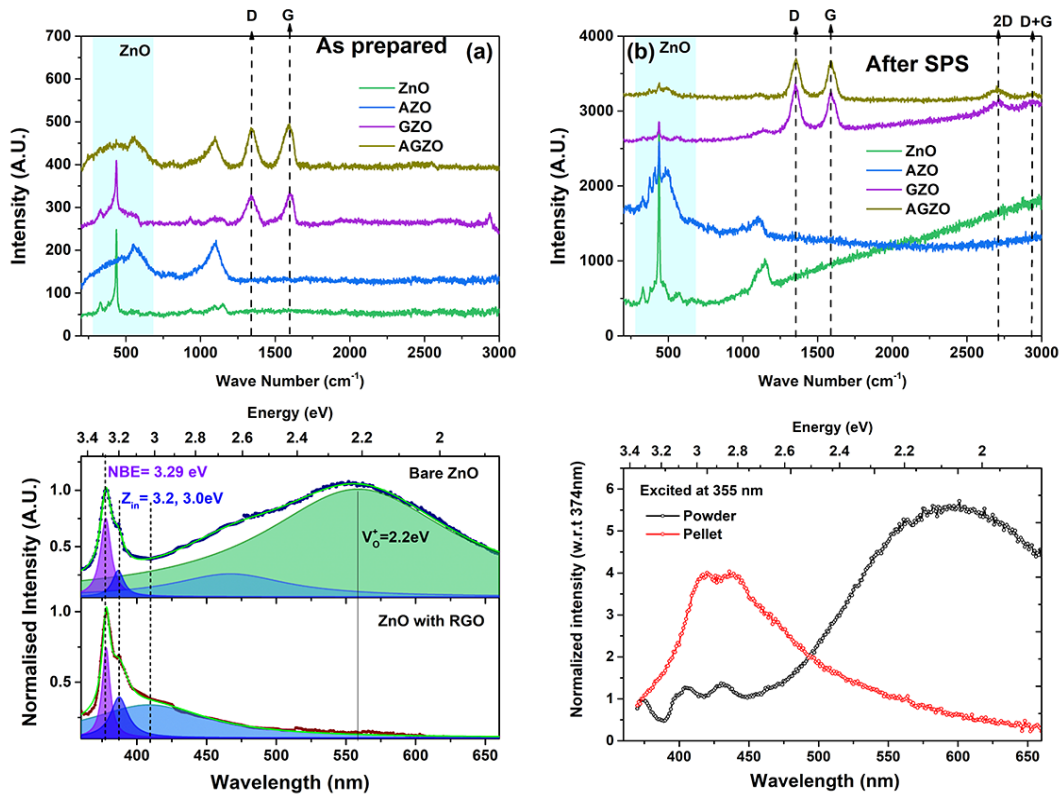

*Figure 6. (a) and (b) Raman spectra of all the powder samples and pellets respectively. (c) PL spectra comparison of ZO and GZO and (d) Before and after pelletization of ZO.*

The photoluminescence (PL) spectra of ZnO QDs with and without RGO shells was studied. As seen from the Fig 6(c), the bare ZnO shows Near Band Edge (NBE) peak at 3.29 eV. In addition, a large broad peak in the visible region is seen which is ascribed to emission from oxygen vacancies ($V_o^+$) levels. Besides, there is a notable intensity at about 3 eV, depicting a significant contribution from Zn interstitial to Valence Band Minimum (VBM) transitions observed at 3.2, 3 and 2.8 eV[37, 38]. However, in case of RGO encapsulated ZnO QDs (GZO), the NBE and Zn interstitial level emissions are observed to be retained. Nonetheless, the peaks attributed to oxygen vacancies has been found to be suppressed drastically. For the given small size (~20 nm), the high surface to volume ratio justifies the presence of large number of surface oxygen vacancies. Moreover, these vacancies also arise due to dangling bonds on surface which are now saturated with interaction with RGO. Thus, RGO encapsulation could suppress the surface oxygen vacancies as reported by Son et.al.[27].

Fig. 6(d) shows the PL spectra comparison of ZnO QDs before and after densification. The excitation wavelength 355 nm was used to see the effect of sintering in high pressure and temperature[37]. Here, the densification is seen to result in oxygen vacancies quenching, again due absence of porosity and texture effect seen from TEM image of sintered sample (Fig 3(d)). Nevertheless, the PL spectra of as prepared powder sample shows the NBE emission peak at 374 nm (3.31 eV), which is suppressed in the ZnO sample after densification. On the other hand, the densified sample shows a strong and broad emission band in the violet (400 nm) and blue regions (500 nm), corresponding to interstitial defect states which are in excess. These defects, which become active in the visible region, arise due to SPS process parameters, primarily the high temperature and the inert (i.e. oxygen-deficient) ambient of the process. Several literature reports show high concentration of zinc interstitials and oxygen vacancies in vacuum treated samples[15, 19, 20, 37].

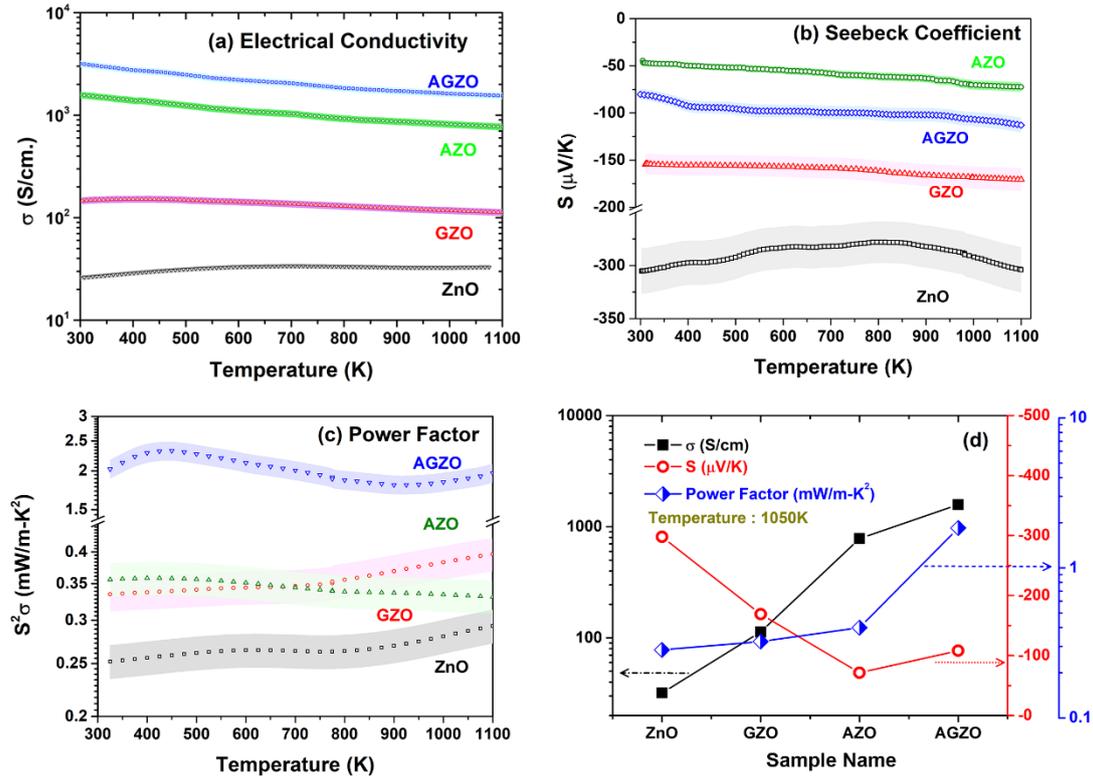

*Figure 7. Temperature dependent (a) electrical conductivity (σ) (b) seebeck coefficient (S) and (c) total electrical contribution or power factor ($S^2\sigma$) of all the samples (d) comparison of σ, S and PF for all the samples at 1050K.*

The temperature dependence of the electrical conductivity sigma (σ) is presented in Figure 7(a). In the semi log graph σ is plotted as a function of temperature (T). For the undoped ZnO, σ is found to follow the semiconducting nature i.e. the conductivity increases with the temperature up about 700 K and remains practically constant beyond it. However, incorporation of the trace level of Aluminium dopants (1%) increases the σ drastically by about two orders of magnitude. From the single parabolic band model, within the free electron theory approximations, one can write the DC electrical conductivity of the sample as-

$$\sigma = ne\mu \qquad \ldots.. (2),$$

where, $n$ is the charge carrier concentration, $e$ is the charge of electron and $\mu$ is the mobility of charge carriers. Thus, the rise in σ could be due to rise in n or $\mu$ or both. Here, the rise in carrier concentration of the sample due to Aluminum donor ions results the rise in conductivity. Moreover, it exhibits a metal-like temperature dependence. Aluminum when doped in ZnO, becomes a substitutional defect as seen from XRD and Raman studies. This leads to donor levels just below the conduction band. The reduction in σ with increasing temperature is mainly due to increasing electron-phonon scattering at higher temperature. This justifies the observed metallic nature of conductivity[39]. On the other hand, when these quantum dots are encapsulated using RGO, the inter-particle presence of RGO might result in lowering the surface band bending induced barrier height[26]. Consequently, it increases the mobility of the charge carriers, giving a higher conductivity across the particles. The enhanced free movement of the charge carrier is also discussed subsequently in the thermal conductivity (κ) discussion. Thus, as a result of this heterogeneous composite formation, the maximum electrical conductivity is achieved in the AGZO sample. However, one can decipher from the Fig 7(a) that the significant contribution to enhancement in σ arises from Al dopant as compared to RGO inclusion.

The Seebeck coefficient (S), is measured for all the four samples and the data is shown in Fig. 7(b). In the undoped ZnO the highest S value of around -300 µV/K is obtained relative to the other doped/composite samples. This is consistent with the observation of its low electrical conductivity. The sign of seebeck coefficient is determined by the dominant charge carrier types. As seen in Fig 7(b), all the samples are showing negative S values in the entire temperature range studied from room temperature to the 1100 K; indicating the dominant charge carriers as electrons. The S value of rest of the samples are moderate in entire temperature range. Nominally, the S value shows inverse dependence on the carrier concentration. Thus, it was expected that the RGO encapsulated sample would show lowest S than without RGO encapsulated Al doped ZnO. On the contrary, it is observed that the AZO shows the lowest S as compared to AGZO for all the temperatures. From Eq (3)[2] it is clear that this might result from higher effective mass of carriers in AGZO samples.

$$S = \frac{8\pi^2 k_b^2}{3eh^2} m^* T \left(\frac{\pi}{3n}\right)^{2/3} \quad \ldots (3),$$

Where, $k_b$ is the Boltzmann constant, $h$ is Planck's constant, $m^*$ is effective mass of carriers, $T$ is absolute temperature. The large Seebeck effect in Al-RGO-ZnO can also be possible due to energy filtering effect, where carriers get scattered at the boundary of RGO-ZnO interface[23, 40].

Based on these values the total electrical contribution is calculated as the Power Factor (PF, $S^2\sigma$) as shown in Fig. 7(c). It is found for the samples ZnO has ~0.25 mW/m-K and AZO as well as GZO has the PF value ~0.35 mW/m-K. Whereas, it is an order of magnitude higher for the AGZO sample with a value of ~2 mW/m-K. This is an exceptionally high PF value for an oxide material so far. The trend is clearly understood from the values of σ, S and PF plotted for all the four samples measured at 1050 K. The monotonous rise in σ and simultaneously high S value shows unusually high PF for the composite sample.

The temperature dependency of thermal conductivity (κ) of all the samples is measured using thermal diffusivity, specific heat and density. The same has been shown in the Fig. 8(a). The measure specific heat data can be found in Fig S9 in supporting information section.

Overall, the thermal conductivity decreases with the increasing temperature. The thermal conductivity consists of two parts i.e. heat carried by phonons called as lattice contribution ($\kappa_L$) and that of carriers, called as electronic contribution ($\kappa_e$). In order to analyze the thermal conductivity, the electronic contribution was extracted from the total κ using the Widemann-Franz Law given in equation (4).

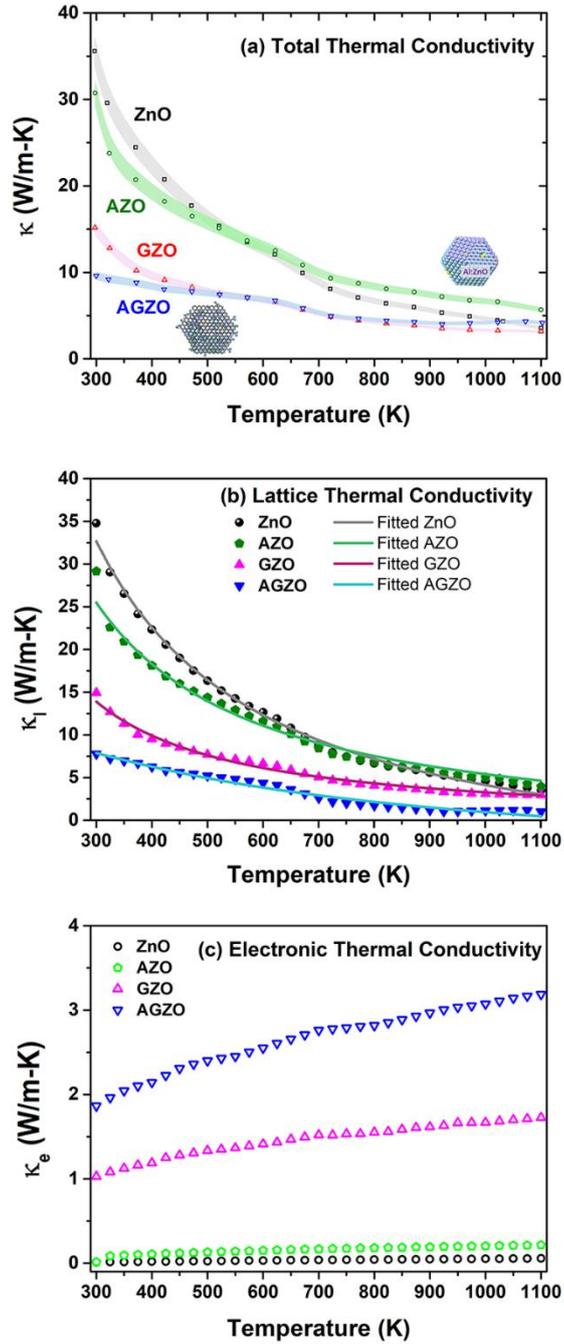

Figure 8. Temperature dependent (a) phononic part or lattice thermal conductivity and the fitted curves (b) electronic part of thermal conductivity and (c) total thermal conductivity of all the samples

$$\kappa_e = L\sigma T \qquad \qquad ...(4)$$

Here, the $L$ is the Lorenz number, which is fitted using the equation (5) reported by Kim et.al.[41] and the $\kappa_L$ is calculated using equation (6).

$$L = 1.5 + e^{-\frac{|S|}{116}} \times 10^{-8} \qquad \ldots (5)$$

$$\kappa_L = \kappa - \kappa_e \qquad \ldots (6)$$

Subsequently, the estimated lattice ($\kappa_L$) and electronic ($\kappa_e$) contributions in the thermal conductivity are calculated, as shown in the Fig. 9(b) and (c) respectively.

The $\kappa_L$ values plotted in Fig 8(b) shows almost the same trend as the total $\kappa$ i.e. sudden reduction upon heating just above room temperature and then saturation at very high temperatures (~1000 K). The practically linear nature at high temperature indicates the dominance of Umklapp scattering process in the thermal transport. This implies that $\kappa_L$ is mainly affected by the phonon-phonon scattering due to inclusion of RGO and point defects created due to the doping and inert ambience from the SPS process. Thus, one may expect that the undoped ZnO should be the most thermally conducting sample than the other three composites. However, at high temperature the AZO and ZnO have nearly same lattice thermal conductivity value. This implies that the contribution due to point defect created by Al doping in AZO is only dominant at low temperature (i.e. 300-600 K). Clearly, the samples having RGO have significantly lower $\kappa_L$ than their RGO deficient counterparts. Thus, the presence of RGO inclusions at the grain boundaries substantially lowers the lattice thermal conductivity contributions as seen from the Fig 8(b). As an effect of this the phonon scattering increased because of the foreign interface, i.e. RGO and it reduced the $\kappa_L$ in the GZO sample. For the AGZO sample not only the phonon-interface scattering dominated but also the point defect effects might have additional scattering centers for phonons. Overall, it results into the drastic suppression of total thermal conductivity in the AGZO sample (nearly one fourth to that of the undoped bare ZnO).

Table 2. The summary of the lattice thermal conductivity analysis.

| Sample code | κ$_L$ at 400K (W/m-K) | | κ$_L$ at 1100K (W/m-K) | | A x 10$^{-43}$ (s$^3$) | B x 10$^{-18}$ (sK$^{-1}$) |
|---|---|---|---|---|---|---|
| | Experimental value | From the Fit | Experimental value | From the Fit | | |
| ZnO | 22.34 | 22.57 | 3.51 | 3.02 | 10.0 ± 0.1 | 32.93 ± 0.55 |
| AZO | 18.11 | 18.30 | 3.95 | 4.64 | 10.28 ± 0.13 | 49.15 ± 0.56 |
| ZG | 9.51 | 9.92 | 2.96 | 2.90 | 9.769 ± 0.12 | 110.72 ± 0.97 |
| AGZO | 6.24 | 6.28 | 1.00 | 0.48 | 125.47 ± 40 | 36.23 ± 0.12 |

To justify this mechanism involved in reduction of κ$_L$, the experimental data is fitted using the Debye-Callaway model given in Eq (7). This fitting is used to extract the various contributions of phonon scattering. According to this model, the lattice thermal conductivity is given as[9, 16]

$$\kappa_L = \frac{k_b}{2\pi^2 v} \left(\frac{k_b}{\hbar}\right)^3 T^3 \int_0^{\frac{\theta_D}{T}} \frac{x^4 e^x}{\tau_c^{-1}(e^x-1)^2} \, dx + \kappa_2 \quad \ldots (7)$$

where, $x = \frac{\hbar \omega}{k_b T}$, $k_b$ is Boltzmann constant, $\omega$ is phonon frequency, $\theta_D$ is Debye temperature, $\hbar$ is reduced Planck constant, $v$ is group velocity, $\tau_c$ is phonon scattering relaxation time, and κ$_2$ is the correction factor due to the conservative nature of the N process and generally ignored at the numerical calculation. From the Mathisenn's rule the combined relaxation time $\tau_c$ is given by the equation (8) and (9).[42]

$$\tau_c^{-1} = \tau_{PD}^{-1} + \tau_U^{-1} + \tau_{bs}^{-1} \quad \ldots (8)$$

$$\tau_c^{-1} = A\omega^4 + BT\omega^2 e^{\frac{-\theta_D}{3T}} + v/l \qquad \ldots (9),$$

Here, $\tau_{PD}$, $\tau_U$, $\tau_{bs}$ are the relaxation times due to point defect, Umklapp scattering (phonon-phonon) and grain boundary scattering respectively. $l$ is the average grain size of the particle, here it is taken as 80 nm after SPS sintering, as calculated from XRD data. The group velocity, $v$ is taken as 3090 m/s and $\theta_D$ is taken 370 K for ZnO[43]. The Eq. (8) $\tau_{bs}$ is mainly dominated at the low temperature range (~10 K). To fit the data using Eq. (9), ω is replaced in terms of $x$. First, $l$ is fixed at 80 nm, however it was relaxed later to get the best fit and extracted the value of the parameters. The best fit parameters obtained are tabulated in Table 2. Although $\kappa_L$ shows a sudden dip for every sample after about 600 K, the change is relatively steeper in case of AGZO sample. Hence, the experimental value is rather lower in the temperature range 700- 900 K than the fit. Thus there is a significant deviation between the experimental data and fitted curve at high temperature. The observed discrepancy is most likely due to the large number of mobile electrons which scatter the moderate energy phonons. Subsequently, this reduces the phononic contribution to heat conduction such that it is almost negligible as compared to that of the electronic part.

It is seen that the $\kappa_e$ increases with the temperature. Due to substitution of aluminium in ZnO lattice, resulting in enhanced carrier concentration, it is expected that the heat conductance contribution due to excess electrons is also subsequently enhanced. However, the change is not commensurate in Al doped ZnO with respect to undoped one in with and without the RGO encapsulation. In other words, the enhancement in electronic ($\kappa_e$) contributions in RGO encapsulated AZO is much more than that of without encapsulation as compared to their respective undoped counterparts. The RGO induced reduction in the grain boundary barrier which helped the electrons to cross the gain boundaries and resulting in one order enhancement in the $\kappa_e$. As previously written in the electrical contribution section this mechanism of charge transfer increased the electrical conductivity of the RGO included samples (GZO, AZGO). A schematic diagram of this is portrayed in Fig 8(c).

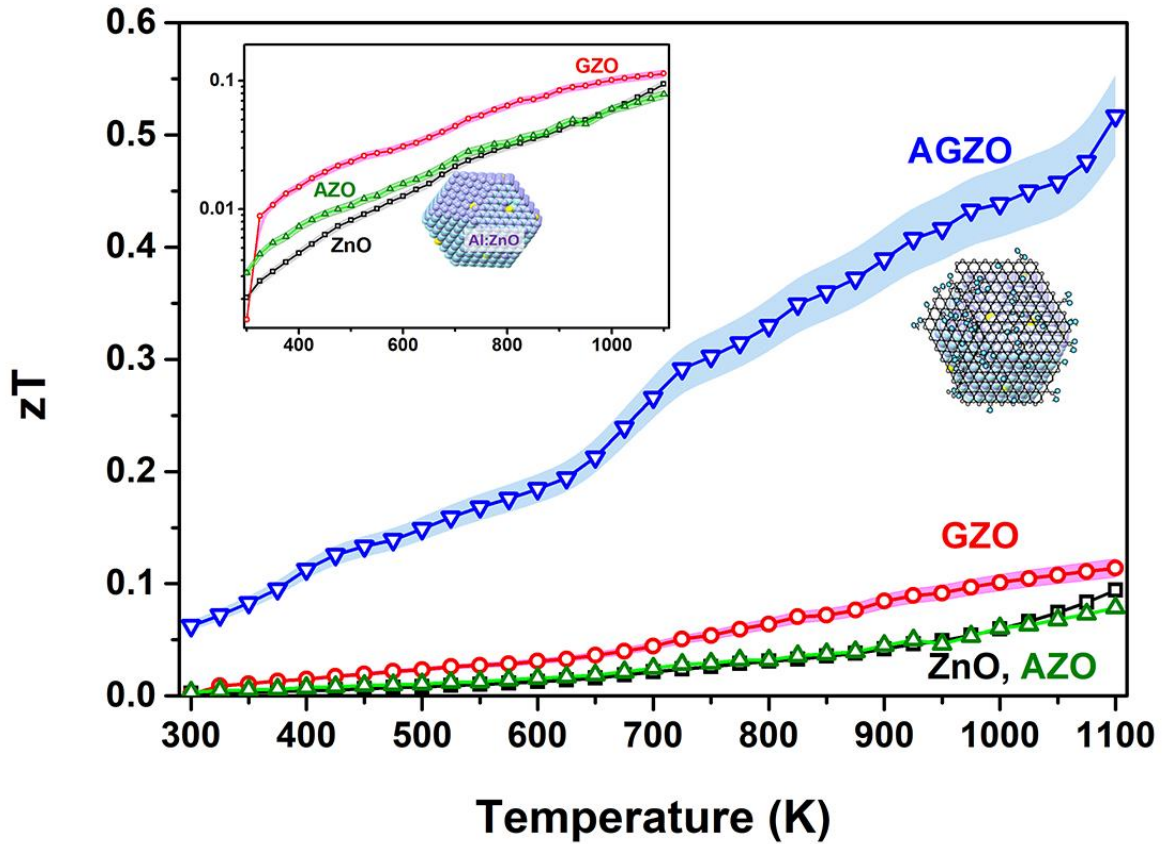

*Figure 9. The dimensionless Figure of merit (zT) of undoped ZnO, Al doped ZnO (AZO), RGO encapsulated ZnO (GZO), RGO encapsulated Al doped ZnO (AGZO) samples as a function of temperature. The inset shows the magnified view of the low zT values of the samples.*

The thermoelectric power generation efficiency of a material is quantified in terms of dimensionless Figure of merit, zT and its variation with temperature (T) is shown in Figure 9. Both the undoped ZnO and Al doped ZnO (AZO) without RGO inclusions show almost identical zT, with weak temperature dependence. The inset shows the same data on semi-log scale. One can see that the Al doping nominally improved zT, only in the low temperature range. The maximum zT value obtained in either case is nearly 0.1 at 1100 K. Here, although the high electrical conductivity and little lower thermal conductivity was obtained for Al doped ZnO, the large seebeck value of undoped ZnO compensated the improvement in other quantities of doped sample. On the other hand, when RGO encapsulation is done on undoped ZnO QDs, there was a notable improvement in the zT, and however, the overall value of maximum zT was not very promising. For the GZO sample the zT barely reached from 0.07 at 300 K to 0.11 at 1100 K. On the other hand, the Al doped as well as RGO encapsulated ZnO QD sample (AGZO) showed a gigantic enhancement of nearly 5 times compared to undoped without RGO encapsulation. The zT value of this sample was found to be 0.52 at 1100 K which is the best value reported for Al doped ZnO system so far to the best of our knowledge. which is only comparable to values obtained for dual doping of Al and Ga[18].

A schematic diagram of the formation of the material and the scattering mechanism are depicted in the Fig. 10.

The solvothermal synthesis method employed, did not involve any high temperature treatments. The modest temperatures of 95 °C during synthesis ensured that the nuclei formed in the solution does not grow and the growth is hindered. Thus, a fine size (~ 20 nm) of ZnO particles were obtained. This is particularly important because the smaller initial size allows to afford a small grain growth which may occur at the sintering stage later[4]. Besides, the solution based method through atomic level mixing also allowed a uniform doping of trace level of Al dopants in the ZnO matrix. Further, introduction of semi-exfoliated GO suspension into this nucleating solution of ZnO, ensured that GO is not only reduced to the RGO sheets but those are wrapped around the ZnO particles effectively. Hence very uniform RGO encapsulation is achieved. The oxygen containing functional groups of RGO could aid in bonding RGO with ZnO, which has a large density of surface oxygen vacancies due to uncompensated bonds as shown in Fig 10(a) and (b).

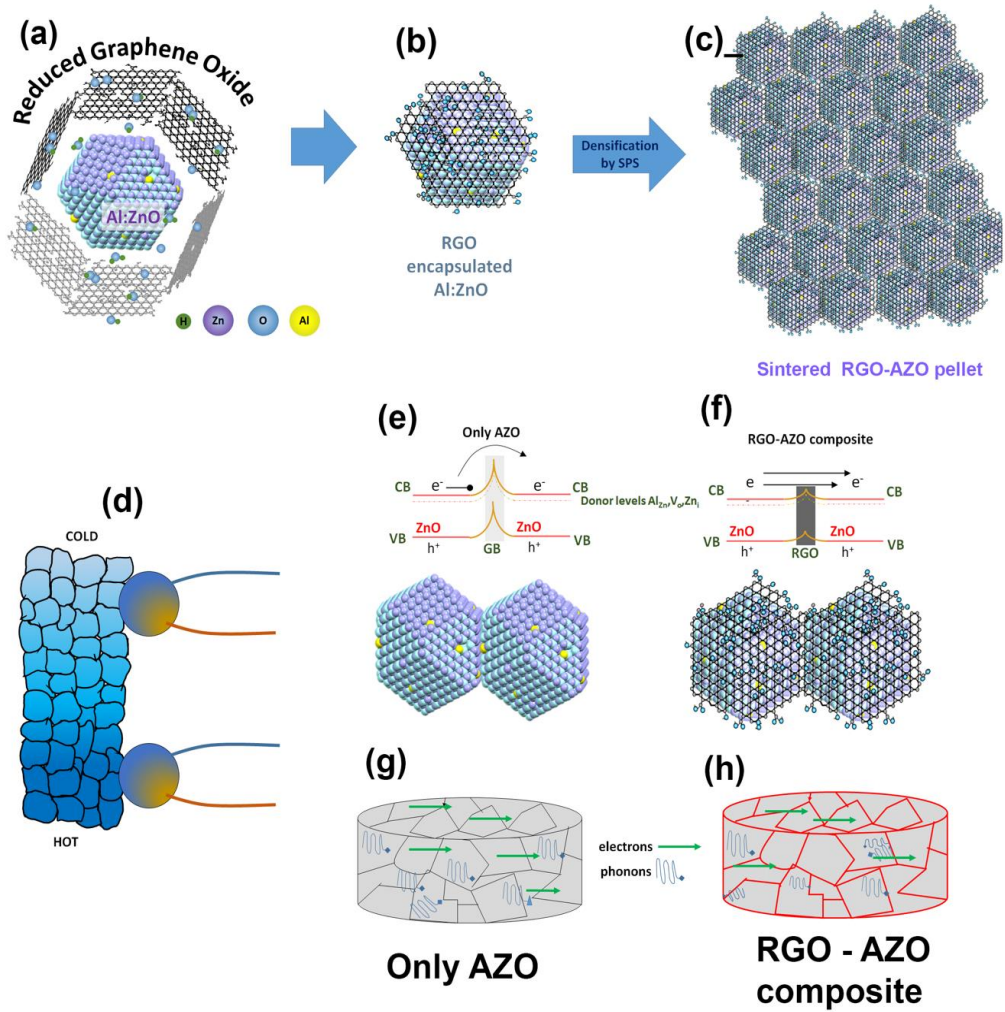

*Figure 10. Schematic diagram of (a) formation of the RGO encapsulated ZnO QDs. (b) an individual RGO-AZO quantum dot, (c) The sintered RGO-AZO composite with dense structure. (d) the schematic showing placement of thermocouples over polycrystalline dense sample. Scattering and transport mechanism in a densified ZnO-RGO composite across grain boundaries (e) without and (c)with RGO encapsulation. (g)& (h) the corresponding scattering of phonons and electrons across in bulk samples.*

The spark plasma sintering method is a non-equilibrium method. i.e. it does not allow the system (sample) to relax in least energy configuration for the supplied thermal energy. Rather, the spark assisted excitation is very short lived and simultaneous uniaxial pressure is applied. As a result, the densification is achieved without any significant grain growth. The temperature of sintering achieved is about 1273 K in inert atmosphere. This high temperature treatment of ZnO in inert (oxygen lean) atmosphere can significantly enhance the oxygen vacancy defects within the ZnO particles. Also the same heat treatment in anaerobic conditions, causes significant reduction of the partially reduced GO. This product is suddenly quenched at room temperature. Hence, as result of SPS sintering, not only high densities are obtained but the grain growth is avoided and it also results in a rise in defects in ZnO (as seen from PL spectra) and reduction of GO to RGO and even graphene-like nature (as seen from Raman and XRD studies).

It is known that the oxygen vacancies in ZnO form the donor levels within the bandgap of ZnO as shown in PL spectra[38]. The higher donor densities, result in high carrier concentration and subsequently high electrical conductivity. Moreover, the conductivity of RGO is also higher than that of GO as the extent of reduction controls the effective bandgap of GO[22, 34, 44]. In turn the resulting product of SPS sintering[45] has very high electrical conductivity due to high mechanical densities as well as high defect concentrations and better conducting RGO interface.

Thus, the overall electrical conductivities of even bare ZnO samples was very high as compared to conventionally sintered samples. This is evident from their resistance of a few mili ohms for bare ZnO and an order of magnitude lesser for AGZO. Besides, clearly the electrical conductivity of Al doped samples was very high than that of without Al doped samples. This is again due to donor levels produced by Al substitutions in Zn site which lies just below the conduction band minima. Thus due to these additional Al donor ions, the electrical conductivity is outstandingly high for an oxide system. Thus these excess electrons contribute to the increased carrier density in AZO.

Normally, most semiconductors have a surface states due to surface adsorption like moisture or oxygen ions. As a result, the surface of ZnO particles and the interface between the two ZnO grains is relatively less conducting compared to the bulk and has an energy barrier as shown in Fig 10(e)[42]. Hence, the conduction in ZnO is mainly limited by the charge transport across the grains. A certain fraction of carriers has sufficient thermal energy to overcome the barrier or tunnel through given the narrow width of barrier. Otherwise a fraction of electrons having energy lesser than that of the barrier are stopped or in other words filtered off as shown in Fig 11(e). This is known as energy filtering process.[23, 40, 46] The electrical conduction maybe improved if this barrier is reduced in height for crossing or made narrower for significant tunneling. Both Al doping as well as RGO encapsulation serve this purpose as Al dopants gives excess carrier concentration while RGO kills the surface states by bonding with the dangling bonds at surface. Thus, the presence of Al dopants and RGO shell reduces surface states and consecutively the barrier between the two grains is substantially lowered (as shown in Fig 10(f)). This further increases the electrical conductivity. Thus, the large enhancement in the electrical conductivity of the composite of RGO encapsulated Al doped ZnO is symbiotic effect of both, RGO capping as well as Al dopants.

On the other hand, the same energy filtering process due to Al doping and RGO shell, is responsible for moderate seebeck coefficient and lowering thermal conductivity[23, 40]. In case of bare ZnO, a good fraction of the free carriers having low energy are blocked by lattice, reducing the effective number of carriers and thus the seebeck coefficient is high. Further Al doping increases carrier concentration and reduces the seebeck coefficient. As noted in the Fig 8(d), although the electrical conductivity of AGZO is the highest, it does not show the lowest seebeck value. Rather it is higher than that of without RGO capped AZO. This can be understood from the nature of seebeck coefficient for composites, especially topological solids, which have conducting edges or surface compared to the bulk. The effective seebeck coefficient of such composites is given by[47],

$$S_{eff} = \frac{G_1 S_1 + G_2 S_2}{G} \qquad \ldots(10)$$

where, $S_1$, $S_2$ and $G_1$, $G_2$ are the seebeck coefficients and conductivities of the constituents' material 1 and material 2 of the composite. $S_{eff}$ and G are the seebeck coefficients and conductivities of the composite. Thus, the total seebeck coefficient of the composite is not only due to AZO (or ZnO) but also has contribution from RGO. In addition, since the measured thermal conductivity of the AGZO composite being lower, justifies that it can sustain large temperature gradients and hence the themoemef is larger than AZO.

The details analysis of thermal conductivity presented in the discussion of Fig 8 points to the fact that in the overall thermal conductivity, the scatterings due to point defects is significant. It is known that the 3D inclusions, precipitates and foreign phases enhances the phonon scattering (as shown in Fig 10(g) and (h)), thereby reducing the lattice thermal conductivity[40]. The cross section of phonon scatterings depends on their mean free path[6]. The point defects and phonon-phonon scattering efficiently scatter only the high frequency phonons, while the grain boundaries scatter the entire spectrum of phonons with equal probability and hence are better heat conduction suppresser.

Hence, considering all these parameter space, it is evident that presence of Al dopant and RGO encapsulation together not only enhance the power factors, but also suppress the lattice contribution of thermal conductivity at high temperatures. This together results in a large enhancement in the thermoelectric figure of merit of ZnO by an order of magnitude. Hence, solvothermal synthesis followed by spark plasma sintering of these particles results in consistently high thermoelectric efficiency for oxides which are more benign materials as compared to the state of the art chalcogenides. The ease of synthesis and lower per unit mass production cost makes them even commercially viable candidates compared to those of chalcogenides.

**CONCLUSIONS**

Undoped and 1% Al doped ZnO quantum dots are synthesized with and without RGO capping of about 1.8 wt% using low temperature solution method. A maximum power factor of ~2 mW m$^{-1}$ K$^{-2}$ is achieved for the AGZO sample at 1100 K which is high for any oxide system. Al doping mainly gave a rise in electrical conductivity and reduced the thermal conductivity, while showing the lower value of Seebeck coefficients. However, the RGO inclusion not only improved overall electrical conductivity but also enhanced seebeck through energy filtering in comparison to bare Al doped ZnO. The simultaneous drop in thermal conductivity of this composite is attributed to the high phonon-phonon scattering, point-defect scattering. The concurrent reduction in the electrical resistivity and thermal conductivity as well as a moderate Seebeck value is a synergic effect which results into the very high zT of 0.52 at 1100 K obtainable by simple processing method. In the subsequent studies, the variation of the RGO capping and doping concentration of Al or even other highly conducting oxides may be tuned to explore further enhancement. Thus, it is demonstrated that the nanostructurization, RGO inclusion, chemical doping through simple solution based method could be utilized as a very effective approach to enhance the figure of merit of environmentally benign ZnO material.

**Supporting Information**

Please see the supporting information section for high resolution XRD, more microscopy images, Raw XPS data on the sintered samples, High resolution Raman spectra, the specific heat data, The STM line profiles.

**AUTHOR INFORMATION**

Corresponding Author

* Email: kbvinayak@iisertvm.ac.in, Phone +91-471-277 8056

**ACKNOWLEDGEMENTS**

The Authors are thankful to Science and Engineering Research Board (SERB) Government of India for the research grant (Number EEQ/2018/000769) which supported this work. Authors are also thankful to Professor Tsunehiro Takeuchi from Energy Materials Laboratory of Toyota Technological institute for providing high temperature thermoelectric properties measurements facilities.

**CONFLICT OF INTEREST**

The authors declare no conflict of interest.

# REFERENCES

1. Bahrami A, Schierning G, Nielsch K. Waste Recycling in Thermoelectric Materials. *Advanced Energy Materials* 2020, **10**(19): 1904159.

2. Shi X-L, Zou J, Chen Z-G. Advanced Thermoelectric Design: From Materials and Structures to Devices. *Chemical Reviews* 2020, **120**(15): 7399-7515.

3. He J, Tritt TM. Advances in thermoelectric materials research: Looking back and moving forward. *Science* 2017, **357**(6358): eaak9997.

4. Ortega S, Ibáñez M, Liu Y, Zhang Y, Kovalenko MV, Cadavid D, *et al.* Bottom-up engineering of thermoelectric nanomaterials and devices from solution-processed nanoparticle building blocks. *Chemical Society Reviews* 2017, **46**(12): 3510-3528.

5. Kim HS, Liu W, Chen G, Chu C-W, Ren Z. Relationship between thermoelectric figure of merit and energy conversion efficiency. *Proceedings of the National Academy of Sciences* 2015, **112**(27): 8205-8210.

6. Yang J, Xi L, Qiu W, Wu L, Shi X, Chen L, *et al.* On the tuning of electrical and thermal transport in thermoelectrics: an integrated theory–experiment perspective. *npj Computational Materials* 2016, **2**(1): 15015.

7. Tan G, Zhao L-D, Kanatzidis MG. Rationally Designing High-Performance Bulk Thermoelectric Materials. *Chemical Reviews* 2016, **116**(19): 12123-12149.

8. Snyder GJ, Toberer ES. Complex thermoelectric materials. *Nature Materials* 2008, **7**(2): 105-114.

9. Banerjee R, Chatterjee S, Ranjan M, Bhattacharya T, Mukherjee S, Jana SS, *et al.* High-Entropy Perovskites: An Emergent Class of Oxide Thermoelectrics with Ultralow Thermal Conductivity. *ACS Sustainable Chemistry & Engineering* 2020, **8**(46): 17022-17032.

10. Assadi MHN, Gutiérrez Moreno JJ, Fronzi M. High-Performance Thermoelectric Oxides Based on Spinel Structure. *ACS Applied Energy Materials* 2020, **3**(6): 5666-5674.

11. Romo-De-La-Cruz C-O, Chen Y, Liang L, Williams M, Song X. Thermoelectric Oxide Ceramics Outperforming Single Crystals Enabled by Dopant Segregations. *Chemistry of Materials* 2020, **32**(22): 9730-9739.

12. Ohta H, Sugiura K, Koumoto K. Recent Progress in Oxide Thermoelectric Materials: p-Type Ca3Co4O9 and n-Type SrTiO3−. *Inorganic Chemistry* 2008, **47**(19)**:** 8429-8436.

13. Nam WH, Lim YS, Choi S-M, Seo W-S, Lee JY. High-temperature charge transport and thermoelectric properties of a degenerately Al-doped ZnO nanocomposite. *Journal of Materials Chemistry* 2012, **22**(29)**:** 14633-14638.

14. Biswas K, He J, Blum ID, Wu C-I, Hogan TP, Seidman DN*, et al.* High-performance bulk thermoelectrics with all-scale hierarchical architectures. *Nature* 2012, **489**(7416)**:** 414-418.

15. Chen D, Zhao Y, Chen Y, Wang B, Chen H, Zhou J*, et al.* One-Step Chemical Synthesis of ZnO/Graphene Oxide Molecular Hybrids for High-Temperature Thermoelectric Applications. *ACS Applied Materials & Interfaces* 2015, **7**(5)**:** 3224-3230.

16. Jood P, Mehta RJ, Zhang Y, Peleckis G, Wang X, Siegel RW*, et al.* Al-Doped Zinc Oxide Nanocomposites with Enhanced Thermoelectric Properties. *Nano Letters* 2011, **11**(10)**:** 4337-4342.

17. Orlinskii SB, Schmidt J, Baranov PG, Lorrmann V, Riedel I, Rauh D*, et al.* Identification of shallow Al donors in Al-doped ZnO nanocrystals: EPR and ENDOR spectroscopy. *Physical Review B* 2008, **77**(11)**:** 115334.

18. Ohtaki M, Araki K, Yamamoto K. High Thermoelectric Performance of Dually Doped ZnO Ceramics. *Journal of Electronic Materials* 2009, **38**(7)**:** 1234-1238.

19. Han L, Van Nong N, Zhang W, Hung LT, Holgate T, Tashiro K*, et al.* Effects of morphology on the thermoelectric properties of Al-doped ZnO. *RSC Advances* 2014, **4**(24)**:** 12353-12361.

20. Zhang D-B, Li H-Z, Zhang B-P, Liang D-d, Xia M. Hybrid-structured ZnO thermoelectric materials with high carrier mobility and reduced thermal conductivity. *RSC Advances* 2017, **7**(18)**:** 10855-10864.

21. Feng X, Fan Y, Nomura N, Kikuchi K, Wang L, Jiang W*, et al.* Graphene promoted oxygen vacancies in perovskite for enhanced thermoelectric properties. *Carbon* 2017, **112:** 169-176.

22. Li T, Pickel AD, Yao Y, Chen Y, Zeng Y, Lacey SD*, et al.* Thermoelectric properties and performance of flexible reduced graphene oxide films up to 3,000 K. *Nature Energy* 2018, **3**(2)**:** 148-156.

23. Ou C, Hou J, Wei T-R, Jiang B, Jiao S, Li J-F, *et al.* High thermoelectric performance of all-oxide heterostructures with carrier double-barrier filtering effect. *NPG Asia Materials* 2015, **7**(5): e182-e182.

24. Nam WH, Lim YS, Kim W, Seo HK, Dae KS, Lee S, *et al.* A gigantically increased ratio of electrical to thermal conductivity and synergistically enhanced thermoelectric properties in interface-controlled TiO2–RGO nanocomposites. *Nanoscale* 2017, **9**(23): 7830-7838.

25. Huang L, Lu J, Ma D, Ma C, Zhang B, Wang H, *et al.* Facile in situ solution synthesis of SnSe/rGO nanocomposites with enhanced thermoelectric performance. *Journal of Materials Chemistry A* 2020, **8**(3): 1394-1402.

26. Rahman JU, Du NV, Nam WH, Shin WH, Lee KH, Seo W-S, *et al.* Grain Boundary Interfaces Controlled by Reduced Graphene Oxide in Nonstoichiometric SrTiO3-δ Thermoelectrics. *Scientific Reports* 2019, **9**(1): 8624.

27. Son DI, Kwon BW, Park DH, Seo W-S, Yi Y, Angadi B, *et al.* Emissive ZnO–graphene quantum dots for white-light-emitting diodes. *Nature Nanotechnology* 2012, **7**(7): 465-471.

28. Hummers WS, Offeman RE. Preparation of Graphitic Oxide. *Journal of the American Chemical Society* 1958, **80**(6): 1339-1339.

29. Singh S, Hirata K, Byeon D, Matsunaga T, Muthusamy O, Ghodke S, *et al.* Investigation of Thermoelectric Properties of Ag2SxSe1−x (x = 0.0, 0.2 and 0.4). *Journal of Electronic Materials* 2020, **49**(5): 2846-2854.

30. Kamble VB, Bhat SV, Umarji AM. Investigating thermal stability of structural defects and its effect on d[sup 0] ferromagnetism in undoped SnO[sub 2]. *Journal of Applied Physics* 2013, **113**(24): 244307-244307.

31. Yadav S, Nair A, Urs Mb K, Kamble VB. Protonic Titanate Nanotube–Reduced Graphene Oxide Composites for Hydrogen Sensing. *ACS Applied Nano Materials* 2020.

32. Wang M, Jiang L, Kim EJ, Hahn SH. Electronic structure and optical properties of Zn(OH)2: LDA+U calculations and intense yellow luminescence. *RSC Advances* 2015, **5**(106): 87496-87503.

33. Kumar N, Srivastava AK, Patel HS, Gupta BK, Varma GD. Facile Synthesis of ZnO–Reduced Graphene Oxide Nanocomposites for NO2 Gas Sensing Applications. *European Journal of Inorganic Chemistry* 2015, **2015**(11): 1912-1923.

34. Arijit K, Harikrishnan G, Kingshuk B, Amit K, Silva SRP, Joy M. Controlling the macroscopic electrical properties of reduced graphene oxide by nanoscale writing of electronic channels. *Nanotechnology* 2021.

35. Shan G, Xu L, Wang G, Liu Y. Enhanced Raman Scattering of ZnO Quantum Dots on Silver Colloids. *The Journal of Physical Chemistry C* 2007, **111**(8)**:** 3290-3293.

36. Srinatha N, No Y, Kamble VB, Chakravarty S, Suriyamurthy N, Angadi B*, et al.* Effect of RF power on the structural, optical and gas sensing properties of RF-sputtered Al doped ZnO thin films. *RSC Advances* 2016, **6**(12)**:** 9779-9788.

37. Sharma S, Bayikadi R, Swaminathan P. Spark plasma sintering route to synthesize aluminium doped zinc oxide. *RSC Advances* 2016, **6**(89)**:** 86586-86596.

38. Bandopadhyay K, Mitra J. Zn interstitials and O vacancies responsible for n-type ZnO: what do the emission spectra reveal? *RSC Advances* 2015, **5**(30)**:** 23540-23547.

39. Tsubota T, Ohtaki M, Eguchi K, Arai H. Thermoelectric properties of Al-doped ZnO as a promising oxide material for high-temperature thermoelectric conversion. *Journal of Materials Chemistry* 1997, **7**(1)**:** 85-90.

40. Ma Z, Wang C, Lei J, Zhang D, Chen Y, Wang Y*, et al.* Core–shell nanostructures introduce multiple potential barriers to enhance energy filtering for the improvement of the thermoelectric properties of SnTe. *Nanoscale* 2020, **12**(3)**:** 1904-1911.

41. Kim H-S, Gibbs ZM, Tang Y, Wang H, Snyder GJ. Characterization of Lorenz number with Seebeck coefficient measurement. *APL Materials* 2015, **3**(4)**:** 041506.

42. Huang J, Yan P, Liu Y, Xing J, Gu H, Fan Y*, et al.* Simultaneously Breaking the Double Schottky Barrier and Phonon Transport in SrTiO3-Based Thermoelectric Ceramics via Two-Step Reduction. *ACS Applied Materials & Interfaces* 2020, **12**(47)**:** 52721-52730.

43. Liang X, Shen L. Interfacial thermal and electrical transport properties of pristine and nanometer-scale ZnS modified grain boundary in ZnO polycrystals. *Acta Materialia* 2018, **148:** 100-109.

44. Capezza A, Andersson RL, Ström V, Wu Q, Sacchi B, Farris S*, et al.* Preparation and Comparison of Reduced Graphene Oxide and Carbon Nanotubes as Fillers in Conductive Natural Rubber for Flexible Electronics. *ACS Omega* 2019, **4**(2)**:** 3458-3468.

45. Kaiser F, Schmidt M, Grin Y, Veremchuk I. Molybdenum Oxides MoOx: Spark-Plasma Synthesis and Thermoelectric Properties at Elevated Temperature. *Chemistry of Materials* 2020, **32**(5)**:** 2025-2035.

46. Popescu A, Woods LM, Martin J, Nolas GS. Model of transport properties of thermoelectric nanocomposite materials. *Physical Review B* 2009, **79**(20)**:** 205302.

47. Xu Y, Gan Z, Zhang S-C. Enhanced Thermoelectric Performance and Anomalous Seebeck Effects in Topological Insulators. *Physical Review Letters* 2014, **112**(22)**:** 226801.